\documentclass[superscriptaddress,amsmath,amssymb,aps]{revtex4-1}
\usepackage{graphicx}
\usepackage{float}
\usepackage{dcolumn}
\usepackage{bm}
\usepackage{setspace}
\usepackage{subcaption}
\usepackage{mathrsfs}
\usepackage{mathtools}
\usepackage{balance}
\usepackage{helvet}
\usepackage{xfrac}
\usepackage[dvipsnames,svgnames,x11names]{xcolor}

\usepackage[dvipsnames,svgnames,x11names]{xcolor}
\usepackage[marginparwidth=2.5cm]{geometry}
\usepackage[final]{changes}
\definechangesauthor[color=BrickRed]{JMG}
\definechangesauthor[color=Blue]{deleted}
\definechangesauthor[color=Green]{fmt}

\usepackage[USenglish]{babel}
\usepackage{setspace}
\usepackage[most]{tcolorbox}

\begin{document}

\title{Quantifying multi-point ordering in alloys}

\author{James M. Goff}
\email{jmg670@psu.edu}
\affiliation{Department of Materials Science and Engineering, The Pennsylvania State University, University Park, PA 16802, USA}
\author{Bryant Y. Li}
\affiliation{Department of Materials Science and Engineering, The Pennsylvania State University, University Park, PA 16802, USA}
\author{Susan B. Sinnott}
\affiliation{Department of Materials Science and Engineering, The Pennsylvania State University, University Park, PA 16802, USA}
\affiliation{Department of Chemistry, The Pennsylvania State University University Park PA 16802, USA}
\affiliation{Materials Research Institute, The Pennsylvania State University, University Park, PA 16802, USA}
\author{Ismaila Dabo}
\affiliation{Department of Materials Science and Engineering, The Pennsylvania State University, University Park, PA 16802, USA}
\affiliation{Materials Research Institute, The Pennsylvania State University, University Park, PA 16802, USA}
\affiliation{Penn State Institutes of Energy and the Environment, The Pennsylvania State University University Park PA 16802, USA}

\begin{abstract}
\normalfont
A central problem in multicomponent lattice systems is to systematically quantify multi-point ordering. Ordering in such systems is often described in terms of pairs, even though this is not sufficient when three-point and higher-order interactions are included in the Hamiltonian. Current models and parameters for multi-point ordering are often only applicable for very specific cases and/or require approximating a subset of correlated occupational variables on a lattice as being uncorrelated. In this work, a cluster order parameter (ClstOP) is introduced to systematically quantify arbitrary multi-point ordering motifs in substitutional systems through direct calculations of normalized cluster probabilities. These parameters can describe multi-point chemical ordering in crystal systems with multiple sublattices, multiple components, and systems with reduced symmetry. These are defined within and applied to quantify four-point chemical ordering motifs in platinum/palladium alloy nanoparticles that are practical interest to the synthesis of catalytic nanocages. Impacts of chemical ordering on alloy nanocage stability are discussed. It is demonstrated that approximating 4-point probabilities from superpositions of lower order pair probabilities is not sufficient in cases where 3 and 4-body terms are included in the energy expression. Conclusions about the formation mechanisms of nanocages may change significantly when using common pair approximations.
\end{abstract}
\maketitle

\clearpage

\section{Introduction}

Chemical ordering in alloys and other multicomponent crystal systems strongly influences material properties such as mechanical strength, durability, and thermodynamic stability. This includes both the long-range periodic arrangement of elements in ordered alloys such as Cu$_3$Au and the short-range order (SRO) that occurs in solid solution crystal systems over shorter distances.\cite{wolverton_first-principles_1998}  In alloys, SRO can influence the thermodynamic stability as well as mechanical properties; increased SRO in CrCoNi alloys leads to increased hardness.\cite{fisher_strength_1954,zhang_short-range_2020} In semiconductors, the optical and electronic properties are influenced by the chemical ordering.\cite{joannopoulos_theory_1976} At the solid-solution interfaces of alloy catalysts, the adsorption of solution species is correlated with alloy ordering and this influences the electrochemical response.\cite{han_surface_2005} In the case of platinum-based alloy nanoshells applied in hydrogen fuel cells as oxygen reduction catalysts, the chemical SRO and structure of the surface alloy have a strong influence on catalyst durability.\cite{zhang_platinum-based_2015} In these catalytic surface alloys and many other cases, such as in high-entropy oxides, semiconductor crystals with multiple sublattices, and other catalyst systems, the chemical ordering motifs of interest are comprised of multiple points and may span multiple sublattices.\cite{pan_perfect_2020,rost_entropy-stabilized_2015,han_surface_2005} Due to the strong interdependence between SRO and material properties, it is desirable to systematically quantify chemical ordering in substitutional systems.

\begin{figure}[h]
    \centering
    \includegraphics[width=0.6\textwidth]{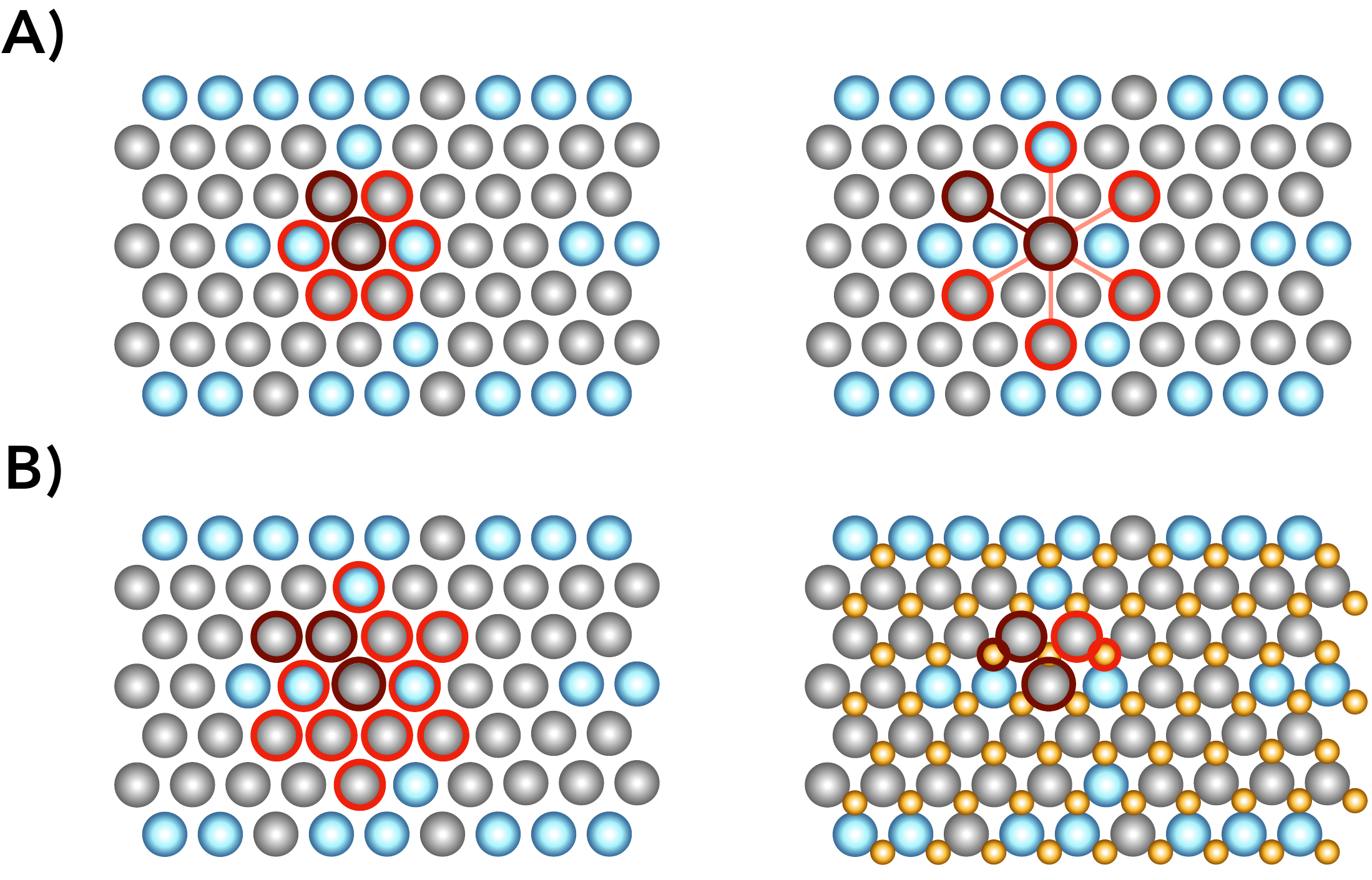}
    \caption{ A) Chemical ordering is often described in terms of pairs (dark red motifs), and how much the chemistry of a given neighbor shell deviates from its nominal stoichiometry on average in the crystal. B) Chemical ordering across multiple shells and/or sublattices can be difficult to quantify, and often requires approximation.}
    \label{fig:alloy_ordering}
\end{figure}

The Warren-Cowley SRO parameters are among the most commonly used descriptors for pair ordering in alloys, both experimentally and theoretically.\cite{cowley_xray_1950,cowley_approximate_1950,mirebeau_first_1984,mohri_first-principles_1991,schonfeld_microstructure_1996,fernandez-caballero_short-range_2017} These parameters can be written for binary systems as 
\begin{equation}
    \gamma^{pq}_m = 1-\frac{P(q|p)_m}{c_q}.
    \label{wc}
\end{equation}
The parameter is given in terms of the conditional probability that atom $p$ is at a site with atom $q$ in some neighbor shell around it, labeled by $m$ (Fig.~\ref{fig:alloy_ordering}A). These probabilities, which can be obtained by inversion of pair correlations, are then divided by the concentration $c_q$. In a random alloy the pair parameter is 0, when $\gamma^{pq} >0$ there is a tendency of $p$-$q$ ordering, and when $\gamma^{pq} <0$ there is a tendency of $p$-$p$ and $q$-$q$ pair ordering. While the description of chemical SRO in terms of pair ordering is useful in many cases; pairs alone do not completely describe a substitutional system. For example in Fig.~\ref{fig:alloy_ordering}B, the Warren-Cowley parameters do not describe the ordering motif where a blue atom occupies a site adjacent to a grey-blue pair,\added[id=JMG,comment={Referee 1 comment 3 -"ordering" used in place of correlation where applicable}]{ a three-point ordering}. \added[id=JMG,comment={Referee 1 comment 3}]{It also cannot describe a three-point ordering} between sublattices in an alloy oxide. While the high-order (3+ body) correlations of this sort are often less significant than pair correlations, all of the $n$-body correlations are needed to completely describe a system.\cite{sanchez_generalized_1984} Many alloys and substitutional crystal systems can be represented with an Ising-like Hamiltonian that depends on chemical occupation variables of sites in the lattice. \added[id=JMG,comment={Referee 1 comment 3 - more explicitly highlighting the case of cluster expansions where any $n$-point correlation is related to an energy contribution in the Hamiltonian (in theory). }]{In such models, it has been proven that correlations up to the order of the interactions in the Hamiltonian (\textit{e.g.} 3-point correlations if 3-body interactions are included)} are required to completely describe all other correlations in the system.\cite{nicholson_relationship_2006} Neglecting many-body correlations, 3+ atom sites, can lead to poor predictions of material properties, as in the inverse Monte Carlo method.\cite{schweika_short-range_1989,laks1992efficient,wolverton_invertible_1997} Incorporation of high-order correlations and associated SRO into models and analysis of substitutional systems would be beneficial, but it is often dismissed due to challenges in obtaining of the multi-point probabilities both experimentally and theoretically.

A number of other developments have been made to extend the Warren-Cowley parameters beyond their typical application to pair-ordering in AB alloys, including the extension of the Warren-Cowley parameters to systems with more than two components.\cite{fontaine_number_1971,ceguerra_short-range_2012} \added[id=JMG,comment={Referee 1 comment 3}]{Work by Clapp showed that some multi-point correlations can be obtained from lower order ones through the Kirkwood superposition (multiplication of pair probabilities).}\cite{clapp_correlation_1966,clapp_exact_1967,clapp_theoretical_1969,kirkwood_statistical_1935} In special cases such as linear binary chains or in equimolar A-B alloys, some of the high-order correlations may be exactly expressed in terms of lower order pair and single-site correlations. In general this is approximate and not suitable for systems with highly correlated lattice occupations or with strong multi-point interactions.\cite{clapp_exact_1967,nicholson_relationship_2006} Similar approximations have been made by Shirley and Wilkins, reconstructing multi-point correlations from pair combinations contained in the motif.\cite{shirley_many-site_1972,shirley_disordered_1977} This method still approximates the occupations of correlated lattice sites contained in the motif as combinations of pairs that occur independently from one another. Its main utility is at the order/disorder transition temperature. Definitions of multi-point order parameters were included in Shirley and Wilkins, but these suffer from the deficiencies associated with the approximated multi-point correlations used to define them. It was proven in Nicholson et al. and demonstrated in other works, that such methods only work in cases where interactions beyond pairs are negligible.\cite{clapp_atomic_1971} When only pair interactions are significant, obtaining multi-point orderings from pair correlations can be highly successful.\cite{gratias_cvm_1985,de_ridder_iterative_1975} Some three-point and four-point ordering parameters have been defined and used for stochastic generation of 2D substitutional lattices possessing high-order correlations.\cite{welberry_solution_1977,welberry1986multi} Methods such as the geometrical locus method that quantify the ordering of derivative structures are currently limited to specific crystal systems and motifs.\cite{brunel_determination_1972,sauvage_prediction_1974,dyck_diffraction_1980} Exact quantification of general multi-point orderings is still needed for substitutional with multiple components and between sublattices. Approximating these from low order correlations is desirable for connection to experimental SRO intensities, but as we show within, does not apply well for all systems with many-body interactions above pairs.

The extraction of three-point and higher correlations from crystals in x-ray experiments is still an active area of study.\cite{lehmkuhler_detecting_2014,pedrini_two-dimensional_2013, pedrini_model-independent_2017,lemieux_investigating_1999} Impressive strides have been made in energy-resolved scanning tunneling electron microscopy to directly measure SRO domains in alloys, but atomic-level chemical ordering across multiple points in alloy systems is still challenging to quantify.\cite{zhang_direct_2019,zhang_short-range_2020} Simulation and theory could be used to directly evaluate multi-point chemical ordering to support experimental findings, but it can be challenging to obtain meaningful statistics in substitutional/alloy systems with many degrees of freedom. The notions of statistical efficiency and accuracy need to be addressed as descriptors of chemical ordering, such as the Warren-Cowley parameter, are extended to multi-point motifs. This was partly addressed by the work of de Fontaine when the order parameters were recast as normalized pair probabilities, and the number of independent pair parameters were defined for systems with arbitrary numbers of components. We aim to extend the description of normalized probabilities to multi-point ordering in alloys. In this work, ClstOPs are introduced for systematically quantifying multi-point ordering in multicomponent crystals through direct measurements of normalized cluster probabilities.

\section{Defining the cluster order parameters}
\subsection{Order parameters on a single sublattice}

We define a set of order parameters to quantify arbitrary multi-point chemical ordering. Like the Warren-Cowley parameters and the current three- and four-point parameters in the literature, the new set of parameters should be 0 for the disordered phase. To begin, an occupation variable representation of a single substitutional lattice is adopted, much like that for Ising or cluster expansion models. The occupations of sites on a lattice are designated by a collection of variables as a occupation/spin vector,
\begin{equation}
    \vec{\sigma} = \{ \sigma_1 , \sigma_2, \sigma_3 \; ... \; \sigma_N \}
    \label{conf_vec}
\end{equation}
A spin variable $\sigma_i$ is assigned to each lattice site of an $N$-site crystal. The spin variables are integers that take on the values:
\begin{equation}
    \sigma_i =
    \begin{cases}
     -m, -m +1,\, ... -1, 1, ... m+1, m \; : \;\;\;\;\;\;\;  m=d/2\\
     -m, -m +1,\, ... -1, 0, 1, ... m+1, m \; : \;\;\;\;  m=(d-1)/2
    \end{cases}
    \label{variables}
\end{equation}
where the first case occurs when the compositional degrees of freedom for a lattice site, $d$, are even and the second case when the degrees of freedom are odd. In a binary alloy containing two species A and B, the spin variables can take the values $\sigma_i = \{-1,1\}$ corresponding to A and B, respectively.

\begin{figure}[h]
    \centering
    \includegraphics[width=0.6\textwidth]{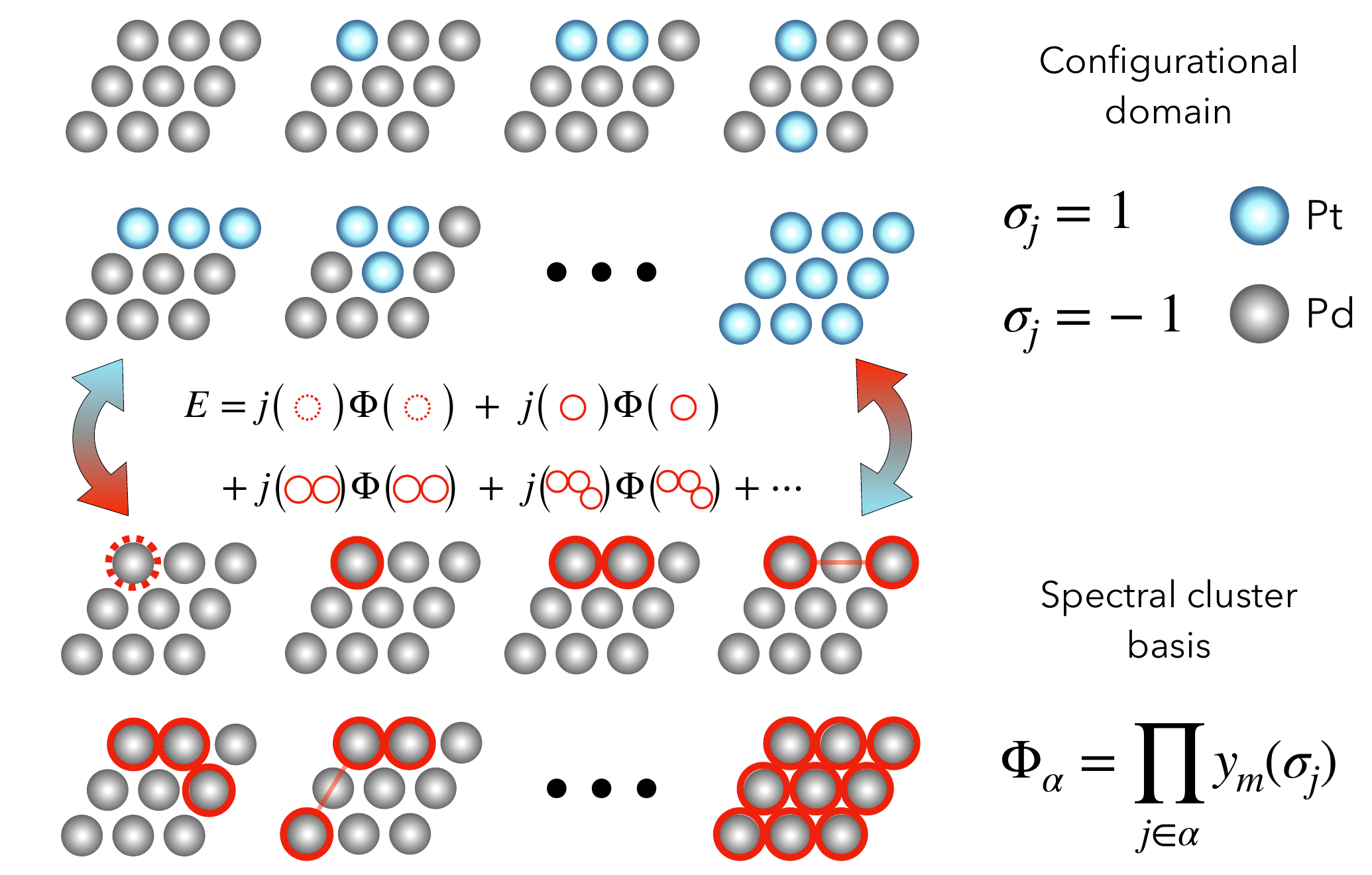}
    \caption{A spectral expansion of the energy is enabled by defining a complete basis over the entire domain of alloy configurations. The alloy configuration is given by a collection of spin variables that specify the occupation of the lattice sites. The basis functions are defined as all possible products the spin variables (or appropriate site basis function to maintain orthogonality conditions). For a cluster expansion in this basis, there are terms corresponding single sites and pairs (of different ranges) similar to an Ising model, but also all high-order terms such as triplets and beyond. }
    \label{fig:ce_fig}
\end{figure}

We are interested in developing an order parameter for arbitrary collections of lattice points in a crystal. The complete, orthonormal set of basis functions defined for the discrete variable space in Eq. \eqref{conf_vec} are a convenient starting point.\cite{sanchez_generalized_1984} \added[id=JMG,comment={Clarification on the cluster expansion formalism is added to address comments 1 and 2 from Referee 1.}]{The cluster basis, spin products, are defined over the entire configurational domain of an alloy crystal, Fig.~\ref{fig:ce_fig}. A scalar extensive property such as the energy may be represented as a linear expansion in this basis, $E =\sum_\alpha j_\alpha \Phi_\alpha$ where the sum runs over clusters, $\alpha$, $j_\alpha$ are the expansion coefficients associated with the respective cluster basis function, $\bar{\Phi}_\alpha$. The chemical occupations for any combination of lattice sites (\textit{e.g.} two lattice sites, 3 lattice sites, etc.) are related to the measured value a corresponding cluster basis function(s) that are defined for the same lattice points. Therefore the chemical ordering of \textit{any} combination of lattice sites may be described within the cluster expansion formalism, because the cluster basis is complete. This basis and formalism is be defined and the exact relationship shown.}

For a given cluster in the set of cluster basis functions, $\{ \Phi_\alpha(\vec{\sigma}_\alpha) \}$, the measured value of the basis function for an alloy configuration specified by the spin vector, Eq.~$\eqref{conf_vec}$, depends on the spin variables of the sites contained in the cluster, $\vec{\sigma}_\alpha$. Each cluster basis function is given by
\begin{equation}
    \Phi_\alpha = \prod_{j \in \alpha} y_m(\sigma_j),
    \label{spin_prod}
\end{equation}
where the product of particular site basis functions, indexed $m$, is taken over all sites $j$ in the cluster. The cluster basis functions obey the following orthogonality,
\begin{equation}
    \langle \Phi_a (\vec{\sigma})| \Phi_b (\vec{\sigma}) \rangle = \delta_{ab}
    \label{cluster_ortho}
\end{equation}
and completeness 
\begin{equation}
    \sum_\alpha \Phi_\alpha(\vec{\sigma_1}) \Phi_\alpha(\vec{\sigma_2}) =\delta_{12}
    \label{cluster_completeness}
\end{equation}
relationships.\cite{sanchez_generalized_1984} The choice of the site basis in Eq. \eqref{spin_prod} is somewhat arbitrary as long as the corresponding cluster functions obey the completeness and orthogonality conditions, Eqs. \eqref{cluster_ortho} and \eqref{cluster_completeness}. In this work, an appropriate trigonometric basis is used, and the specification of site basis indices, $m$ in Eq. \eqref{spin_prod}, are set by the cluster index, $\alpha$.\cite{angqvist_icet_2019} The set of cluster basis functions includes an empty identity cluster for completeness, single-site, pairs, triplets, quadruplets, and so on as schemed in Fig.~\ref{fig:ce_fig}.

\added[id=JMG,comment={Clarification on the cluster correlations and interaction energies is added to address comments 1 and 2 from Referee 1.}]{In practice for periodic crystals, the symmetry of the crystal imposes constraints on the expansion coefficients. For practical cases, it is more convenient to average the cluster basis functions over the crystal. The average cluster basis functions, A.K.A. cluster correlation functions, are given as }
 \begin{equation}
     \bar{\Phi}_\alpha(\vec{\sigma}) = \frac{1}{m_\alpha N} \sum_{\beta = \alpha}\sum_{p}^{N_p} \Phi_\beta(\vec{\sigma}_\beta(p)),
     \label{avg_bfunc}
 \end{equation}
where the inner sum runs over all distinct locations of the a cluster, $p$, and the outer sum runs over all symmetrically equivalent clusters, $\beta = \alpha$. Per-site correlations are obtained by dividing by the number of lattice sites, $N$, and the number of symmetrically equivalent clusters, $m_\alpha$. It is noted that the number of distinct cluster locations, $N_p$, may differ from the total number of sites, $N$, in crystals with reduced symmetry, such as a 2D surface with a set thickness. This is the case in the example given later. \added[id=JMG,comment={More detailed discussion of the cluster expansion itself is given earlier to more clearly show the connection between the cluster correlations and system energetics}]{A scalar extensive property such as the energy of a substitutional lattices system, $E(\vec{\sigma})$, may be represented per-site as a linear expansion in the cluster correlations.}
\begin{equation}
    E(\vec{\sigma}) = \sum_\alpha m_\alpha J_\alpha \bar{\Phi}(\vec{\sigma}_\alpha),
    \label{ce_energy}
\end{equation}
Here, the sum is performed over all symmetrically distinct clusters, and the coefficients, $J_{\alpha}$, referred to as the effective cluster interactions (ECIs), describe the strength of a interaction averaged over the lattice.

The correlations in Eq. \eqref{avg_bfunc} can be written in terms of occupational pair, triplet, quadruplet etc. probabilities as a weighted average.
 \begin{equation}
     \bar{\Phi}_\alpha(\vec{\sigma}) =\frac{N_p}{N} \sum_{\vec{\sigma}_\alpha} \Phi_\alpha(\vec{\sigma}_\alpha) \bar{P}(\vec{\sigma}_\alpha)
     \label{bfunc_prob}
 \end{equation}
Here, the sum now runs over all distinct occupations of the cluster multiplied by the respective probability of finding any symmetrically equivalent cluster with that specific occupation in the crystal, $\bar{P}(\vec{\sigma}_\alpha)$.
These probabilities, referred to as cluster probabilities, can be defined as:
\begin{equation}
    \bar{P}(\vec{\sigma}_\alpha) = \frac{1}{m_\alpha \,N_p}\sum_{\beta=\alpha} n(\vec{\sigma}_\beta).
    \label{pmeasure}
\end{equation}
The total number of clusters with the desired occupation is counted at each distinct location, $p$, in the crystal to give $n(\vec{\sigma}_\alpha)$. The counts are summed over all symmetrically equivalent clusters, and divided by the symmetry multiplicity and total number of occurrences of the cluster in the crystal, $N_p$. These probabilities sum to 1:

\begin{equation}
    \sum_{\vec{\sigma}_\alpha} \bar{P}(\vec{\sigma}_\alpha) = 1.
    \label{prob_norm}
\end{equation}
The sum is taken over all possible distinct occupations of the cluster.

\added[id=JMG,comment={Referee 1 comment 3}]{Using the multi-point probabilities defined in Eq. \eqref{pmeasure}, the ClstOP is defined.}
\begin{equation}
    \gamma_\alpha(\vec{\sigma}_\alpha^k) = 1\;-\;\frac{\bar{P}(\vec{\sigma}_\alpha^{k})}{P_{\rm{random}}}
    \label{clst_order}
\end{equation}
where the order parameter, $\gamma_\alpha$ is given in terms of the average probability of finding a cluster with a desired occupation, $(\vec{\sigma}_\alpha^{k})$ in the crystal that is normalized by the probability of the cluster forming with the desired occupation in a random alloy, $P_{\rm{random}}$.
In the random alloy the probability of a site being occupied by a specific species, $P(\sigma_i)$, is given by the atomic concentration of that species $C_i$; the probability in the denominator is the product of the site probabilities for a given cluster occupation. For example, $P_{\rm{random}}$ in an AB alloy for a three-point occupation of $(\vec{\sigma}_\alpha^{k}) = [-1,-1,-1]$ corresponding to (AAA) is given by $C_A C_A C_A$. It can be directly shown that by selecting a pair cluster, $\alpha \, \in \, \{\rm{pairs}\}$ the \added[id=JMG,comment={Referee 1 comment 3}]{ClstOP} reduces to the binary Warren-Cowley SRO parameters. 

The analysis of the ClstOPs is similar to that for the Warren-Cowley pair parameters.\cite{cowley_approximate_1950} When the ClstOP is zero, then the cluster shape with the specified occupation occurs as frequently as it would in a random alloy. When $\gamma_\alpha > 0$ then the cluster with the specified occupation is found less often than in a random alloy of the same composition. Finally, when $\gamma_\alpha < 0$ then the cluster with desired occupation occurs more frequently than in a random alloy. \added[id=JMG,comment={Cluster expansion in terms of the cluster order parameter is defined to more clearly show the connection between the chemical ordering and system energetics.}]{The cluster expansion of the energy in Eq. \eqref{ce_energy} may be written in terms of the ClstOPs. } Using Eqs. \eqref{bfunc_prob} and \eqref{ce_energy}, the cluster expansion of the energy may be written as,
\begin{equation}
    E(\vec{\sigma}) = \frac{N_p}{N}\sum_\alpha m_\alpha J_\alpha \sum_{\vec{\sigma}_\alpha} \Phi_\alpha(\vec{\sigma}_\alpha) P_{\rm{random}}[1-\gamma_\alpha(\vec{\sigma}_\alpha)],
    \label{ce_clstop}
\end{equation}
where the outer sum runs over all symmetrically distinct clusters, and the inner sum runs over the chemical labelings of that cluster. In this equation, the cluster correlations, $\bar{\Phi}_\alpha$, have been rewritten in terms of ClstOPs rather than the cluster probabilities as in Eq. \eqref{bfunc_prob}. It is noted that the inner sum runs over the same cluster labelings and sites for multicomponent cluster functions belonging to the same orbit; the different site basis functions just give a different value of $\Phi_\alpha(\vec{\sigma}_\alpha)$.\cite{sanchez_generalized_1984} What can be inferred from Eq. \eqref{ce_clstop} is that sums of the chemical ordering parameters scaled by the evaluated basis function for a given cluster labeling, $\Phi_\alpha(\vec{\sigma}_\alpha)$, determines the energy contribution from a given cluster. If the expansion coefficient associated with a cluster correlation, $\bar{\Phi}_\alpha$, is large, the influence of the associated chemical orderings will have more of an impact on the system energetics. The relationship between the chemical ordering and the energy is not linear in general; the value of the ClstOP, $\gamma_\alpha (\vec{\sigma}_\alpha^k)$, and all other ClstOPs associated with the possible occupations of the cluster, $\gamma_\alpha (\vec{\sigma}_\alpha^{\kappa \ne k })$, are constrained dependently by the composition of the system. 

\added[id=JMG,comment={Clarified discussion of consequences long vs short range correlations on the energy and added discussion on how long & short range orderings could be used to extract energetic contributions.}]{One benefit of the completeness of the cluster basis is that the energy contribution from any correlation may be calculated, and through Eq. \eqref{ce_clstop}, the energy contributions from any set of associated chemical orderings. These may correspond to long or short-range correlations. Conversely, given a set of chemical orderings for a system can be used to extract expansion coefficients that may yield said chemically ordered structures (inverse Monte Carlo).\cite{gerold_determination_1987} Inverse Monte Carlo performed with ClstOPs could be distinguished from traditional inverse Monte Carlo methods that provide interaction energies from pair ordering alone.} Limiting values of the ClstOPs may be related to long-range or superstructure orderings depending on the cluster and crystal. The limits of the ClstOPs at large separations could be used to generalize long-range order parameters.\cite{cowley_short-range_1965} Special cases of derivative structure orderings and multi-point motifs can be inferred by the geometrical locus method. Recall that the generalized geometrical locus method provided constraints on the pair parameters spanning derivative polyhedra in certain AB crystals (rocksalt, CsCl structure, and SnS structure).\cite{brunel_determination_1972,sauvage_prediction_1974} The ordering of octahedra in the rocksalt lattice can be inferred based on composition, whether or not the octahedra are arranged periodically, the composition of the octahedron, and noting that the octahedra span the crystal. With \added[id=JMG,comment={Referee 1 comment 3}]{ClstOPs} the ordering of octahedra (or other derivative polyhedra) can be directly measured with polyhedral 'cluster'. Defining similar with the cluster probabilities could allow for a generalization of the geometrical locus method to derivative structures beyond polyhedra, and alloys with more than two components.\cite{morgan2017generating} 

\subsection{Order parameters on multiple sublattices}

The \added[id=JMG,comment={Referee 1 comment 3}]{ClstOPs} can be generalized to crystals with multiple sublattices. This is demonstrated in a two sublattice system for example. The occupations of the sites in each respective are designated with a spin vector, $\vec{\sigma}$ and  $\vec{\delta}$ as in Eq. (2) of the main text. A set of cluster basis functions can be assigned to each sublattice, $\{\Phi^1_\alpha(\vec{\sigma})\}$ and $\{\Phi^2_\alpha(\vec{\sigma})\}$ A complete set of composite basis functions can be defined for the multi-lattice crystal by taking the tensor product of the single sublattice clusters spaces.\cite{tepesch_model_1995} Using a similar notation as in Tepesch et al. in 1995 these composite basis functions, indexed by $\theta$, are be denoted as:
\begin{equation}
    \{\Theta_\theta(\vec{\tau}_\theta)\}
    \label{composite_basis}
\end{equation}
where we have defined a single vector containing the spin variables for the entire crystal.
\begin{equation}
    \vec{\tau} = (\sigma_1, \sigma_2, ... \sigma_{N_1} , \delta_1, \delta_2, ... \delta_{N_2} )
\end{equation}
The inter and single-sublattice correlations are given as the expectation value of the composite basis functions over the crystal. These are calculated as
\begin{equation}
     \bar{\Theta}_\theta(\vec{\tau}) = \frac{1}{m_\theta N} \sum_{\phi = \theta}\sum_{c}^{N_c} \Theta_\phi(\vec{\tau}_\phi(c)) .
     \label{avg_composite_bfunc}
\end{equation}
The inner sum runs over all distinct locations of the composite cluster, $c$, and the outer sum now runs over all symmetrically equivalent composite clusters, $\theta=\phi$. The sum of the evaluated composite cluster functions at all of these points is then divided by the total number of sites in the crystal coming from sublattices 1 and 2, $N = (N_1 + N_2)$, multiplied by the number of symmetrically equivalent composite clusters, $m_\theta$.

The correlations in Eq. \eqref{avg_composite_bfunc}, can also be written as weighted averages of cluster basis functions evaluated for specific \textit{cluster} occupations following a similar form as that for a single sublattice.
\begin{equation}
    \bar{\Theta}_\theta(\vec{\tau}) =\frac{N_c}{N} \sum_{\vec{\tau}_\theta} \Theta_\theta(\vec{\tau}_\theta) \bar{P}(\vec{\tau}_\theta)
    \label{composite_bfunc_prob}
\end{equation}
where the sum now runs over all occupations possible in the composite cluster. The multi-point probabilities for the inter-sublattice correlations are defined as
\begin{equation}
    \bar{P}(\vec{\tau}_\theta) = \frac{1}{m_\theta \,N_c}\sum_{\phi=\theta} n(\vec{\tau}_\phi).
    \label{pmeasure_m}
\end{equation}
The counts of composite clusters with a specific inter-sublattice occupation, $\tau_\theta$, are summed for all symmetrically equivalent composite clusters and divided by the total number of occurrences of the composite cluster in the crystal, $N_c$, multiplied by the symmetry multiplicity. With this, the inter-sublattice ClstOP can be defined.
\begin{equation}
    \gamma_\theta = 1\;-\;\frac{\bar{P}(\vec{\tau}_\theta^{d})}{P_{\rm{random}}}
    \label{clst_order_mult}
\end{equation}

The correlations in Eq. \eqref{composite_bfunc_prob} could also be inverted to obtain specific probabilities. In the case of the composite cluster formed by the product of the two single-site basis functions, $\Theta([\sigma_i,\delta_j])=\Phi_{\rm single}^1(\sigma_i)\Phi_{\rm single}^2(\delta_j)$, the inter-sublattice pair probabilities could be extracted. Because correlations beyond pairs can be considered, the ordering of cations about an anion vacancy could be considered in the rocksalt crystal structure for example.\cite{rost_entropy-stabilized_2015} Pair ordering alone would likely show a large tendency of unlike pair ordering between the cation and anion species.\cite{stana_chemical_2016}

\section{Application to Pt-based alloy nanoshell catalyst}
\begin{figure}[h]
    \centering
    \includegraphics[width=\textwidth]{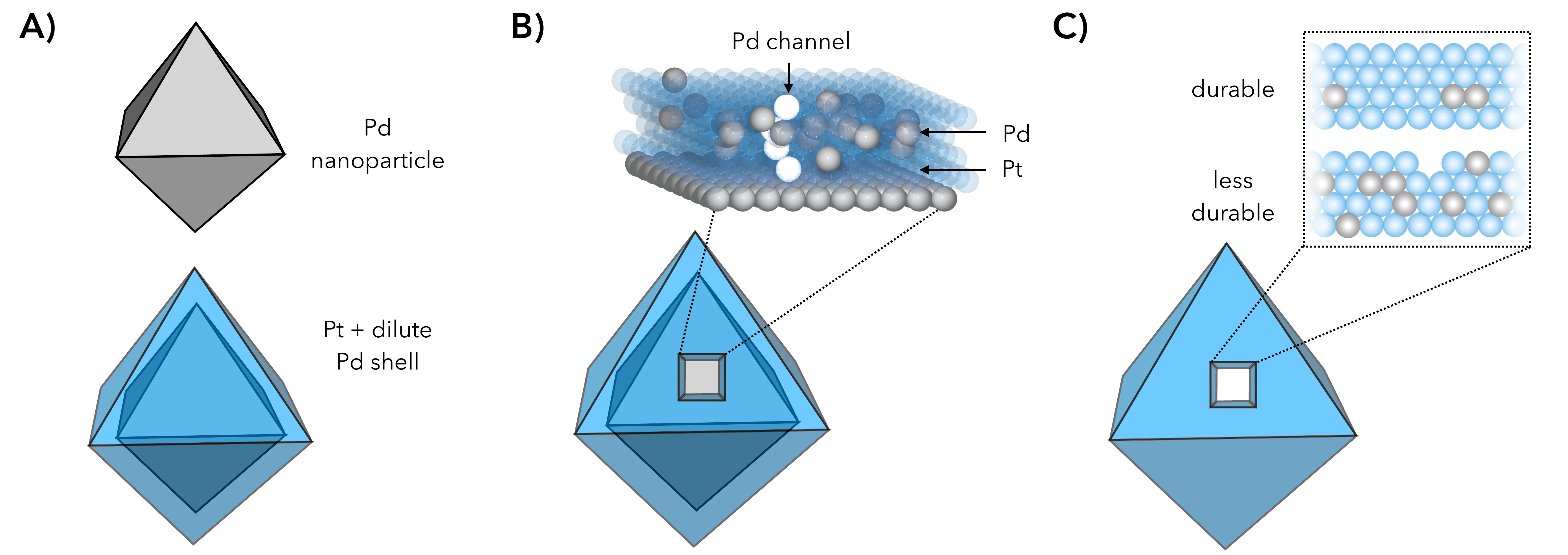}
    \caption{The palladium (111) fcc nanoparticles coated with a few atomic layers of a Pt/Pd alloy, A), prepared in Zhang 2015 rely on specific multi-point chemical ordering motifs, palladium channels that span the surface alloy region B), to generate high-surface area catalyst cages for the oxygen reduction reaction, C). A sufficient amount of Pd is needed to observe the 'channel' order motif in B) to allow for the etching of the palladium core but not so much that it diminishes the catalyst durability.}
    \label{fig:structure}
\end{figure}

The need for an exact description of chemical SRO is highlighted in the case of platinum-based nanoshells in Fig.~\ref{fig:structure}.\cite{zhang_platinum-based_2015} To make the platinum-based nanoshells, thin layers of dilute Pt/Pd alloy are deposited on palladium nanoparticles. The Pd cores are subsequently leached out leaving a highly active, predominantly Pt shell (9.1\% mass Pd). Though the thicknesses of these catalyst shells are as low as 4 atomic layers, this resembles the surface of many other core-shell and alloy interfaces.\cite{mani_dealloyed_2008,alayoglu_rupt_2008,stamenkovic_improved_2007} As evidenced in Zhang et al., the formation of Pd channels that span, or nearly span the deposited surface alloy, allow for subsequent etching of the Pd cores. They also found that excess Pd content decreased mechanical stability of the shells. Nanoshell catalysts with increased durability could potentially be produced by decreasing the Pd content while still allowing for Pd channel formation, Fig.~\ref{fig:structure}B. 

Pair ordering analysis would provide some insights into the Pd channel content in alloy surfaces, but the pair approximations used to quantify the Pd channel occurrence is somewhat arbitrary. The \added[id=JMG,comment={Referee 1 comment 3}]{ClstOPs} were used instead to directly quantify Pd channel content in models of this alloy surface, and was compared to an approximated channel parameter constructed from pair probabilities. The surface of these alloy-coated nanoparticles were modeled using cluster expansions fit with and without a continuum solvent interface above the alloy surface to simulate some effects of the experimental environment.\cite{angqvist_icet_2019,andreussi_revised_2012} In these models, the (111) fcc surface was represented with 4 layers of Pt/Pd alloy on top of four pure palladium layers to represent the palladium core. The occurrence of four-point Pd channels spanning the alloy region determines whether the catalyst can be synthesized and its durability. 

\subsection{Model for the alloy surface}
\subsubsection{Density functional theory calculations}
\begin{figure}[H]
    \centering
    \includegraphics[width=0.7\textwidth]{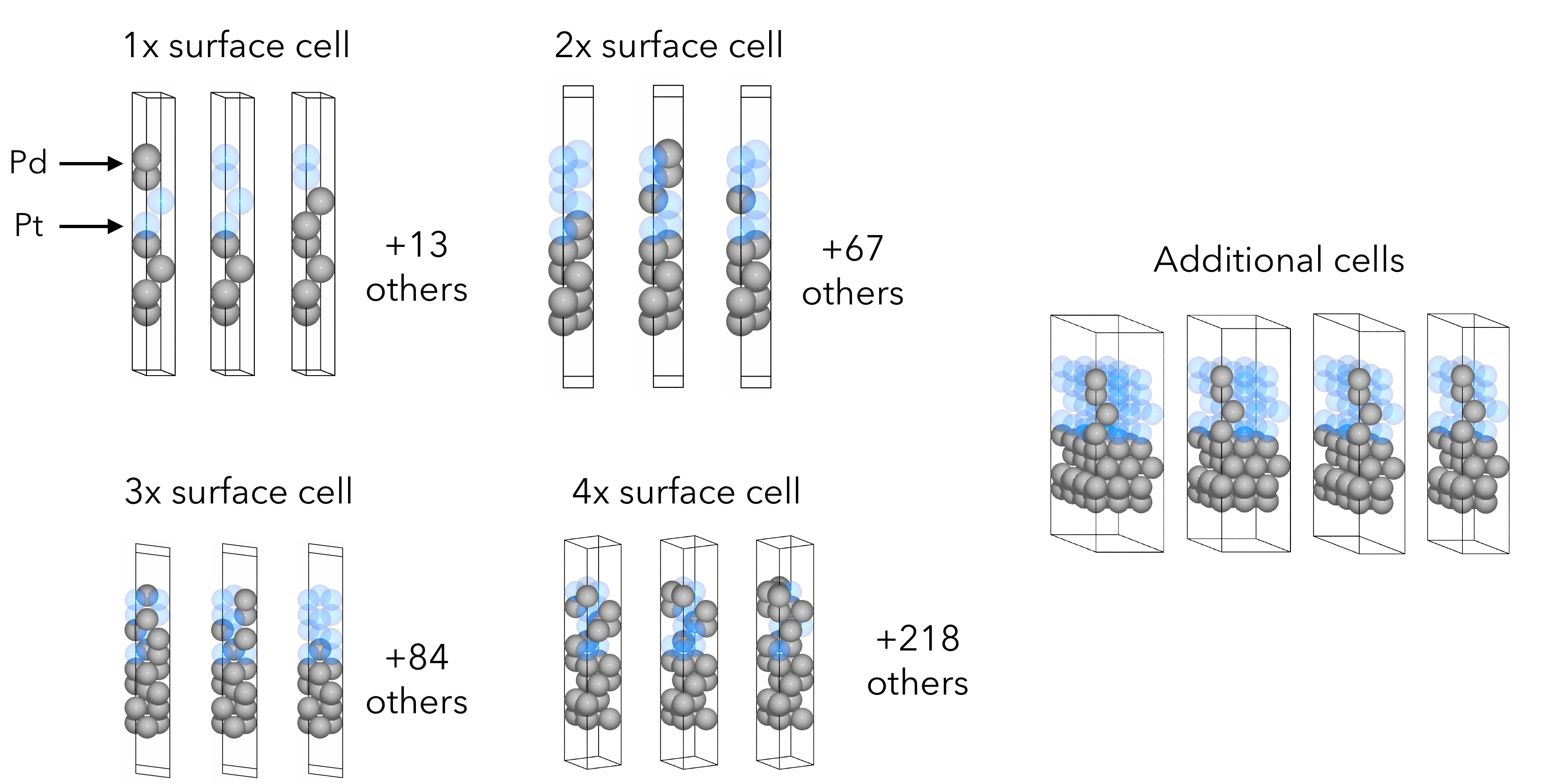}
    \caption{Examples of DFT surface cells used to fit the cluster expansion coefficients (grouped as multiples of the primitive surface cell). All symmetrically distinct configurations in the 1X and 2X cells are used for fitting along with larger randomized surface cells.}
    \label{fig:dft_set}
\end{figure}

A selection of 398 symmetrically unique alloy configurations were used as training data for the cluster expansion. Some examples are provided in Fig.~\ref{fig:dft_set}. The mixing energies of these alloy configurations were calculated using density functional theory. The surface alloy slabs were comprised of 8 layers in total with a vacuum height of 6 ${\rm{AA}}$ on either side. Four of the layers on the bottom were pure Pd to represent the Pd core of the nanoparticles where the top four layers were comprised of both Pt and Pd with varied concentration. The DFT calculations were performed using the Quantum Espresso suite with plane wave basis sets.\cite{giannozzi_quantum_2009} The kinetic energy cutoff for the basis sets was 100 Ry. Norm conserving pseudopotentials from the {\sc PseudoDojo} library were used to represent ion cores.\cite{van_setten_pseudodojo_2018} An approximately uniform distribution of reciprocal Bloch vectors ({\bf k}-points) was used to sample the Brillouin zone across the cells by using the {\bf k}-point density of (11/$m$ $\times$ 11/$n$ $\times$ 1) for an $m$ $\times$ $n$ surface cell. Electronic occupations were smoothed with 0.001 Ry of Marzari-Vanderbilt smearing.\cite{marzari_thermal_1999} The {\bf k}-points, smearing, slab thickness, energy cutoffs, and vacuum height were converged with respect to the Fermi energy of the system, to within 0.05 eV, ensuring that the interfacial dipole was converged. During geometry optimizations of surface alloy slabs the bottom two layers of Pd were fixed at calculated bulk lattice parameters, and total forces were below 25 meV/\AA.
To account for average solvent effects on the surface alloy, the DFT energies for the surface alloy configurations were also calculated in the presence of a continuum solvent via the Self-Consistent Continuum Solvation method.\cite{andreussi_revised_2012} The shape and onset of the dielectric cavity are defined using minimum and maximum charge density cutoffs ($\rho_{\rm{min}}$ = 0.0013 and $\rho_{\rm{max}}$ = 0.01025) the original switching function provided in Andreussi, Dabo, and Marzari was used. The dielectric constant inside the cavity is 1 and switches smoothly to a dielectric constant of 78.3 outside the cavity. Surface tension, pressure, and volume terms were omitted as in Huang et al.\cite{huang_potential-induced_2018} Though the DFT fitting cells in Fig.~\ref{fig:dft_set} are not symmetric, the contributions from solvent on the palladium core side should cancel out when calculating the mixing energies. Models including explicit solvation, adsorption of solution ions, and the etchant will likely show a stronger influence on the surface alloy structure and would help determine solvent conditions suitable for making more durable cages.\cite{han_surface_2005,cao_computationally_2019} 
\subsubsection{Cluster expansions}
\begin{figure}[h]
    \centering
    \includegraphics[width=0.85\textwidth]{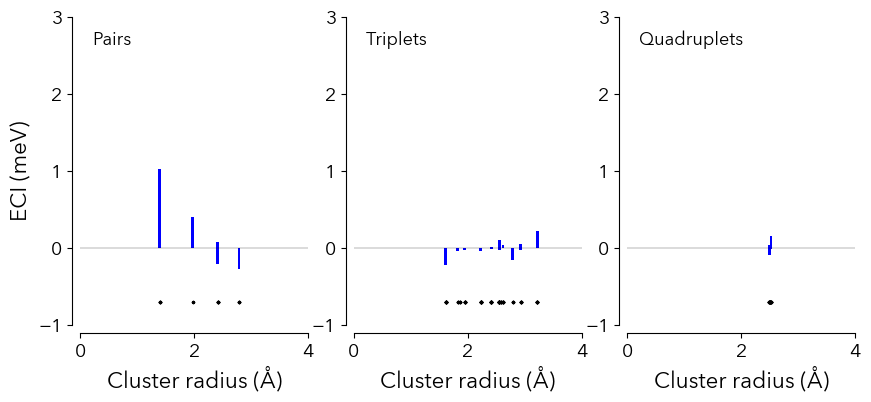}
    \caption{Effective cluster interactions/expansion coefficients for the surface alloy system in contact with the implicit solvent. The blue bars mark the magnitude and sign of the coefficient for pairs, triplet, and quadruplet clusters, order 2, 3, and 4, respectively.}
    \label{fig:ecis}
\end{figure}

The cluster expansion model represents a scalar extensive quantity as a linear expansion in the cluster basis functions of Eq. \eqref{avg_bfunc}, and was obtained using the ICET software package.\cite{angqvist_icet_2019} The per-site mixing enthalpy for the representative surface alloy was expanded as:
\begin{equation}
    \Delta H_{\rm{mix}}(\vec{\sigma}) = \sum_\alpha m_\alpha J_\alpha \bar{\Phi}(\vec{\sigma}_\alpha),
    \label{expansion}
\end{equation}
with the sum taken over all symmetrically distinct clusters up to some maximum size and order. \added[id=JMG,comment={Retrained cluster expansion with smaller set of parameters to reduce overfitting}]{Aside from identity and single site clusters, all pair and triplet clusters with within 3 neighbor shells were included. Additionally, some larger quadruplet clusters are included that span the 4th and 5th neighbor shells. This is done to include the cluster corresponding to the channel shape in Fig. \ref{fig:structure} B) in the energy expression. This resulted in 25 pair, 138 triplet, and 34 quadruplet clusters for a total of 201.} The expansion coefficients were trained against DFT fitting data in Fig. \ref{fig:dft_set} using the Automatic Relevance Determination Regression (ARDR) method implemented in scikit-learn to obtain an optimally sparse set of ECIs and reduce over-fitting yielding \added[id=JMG,comment={Retrained cluster expansion with smaller set of parameters to reduce overfitting}]{68 non-zero ECIs after training}.\cite{pedregosa_scikit-learn_nodate} A weighting function was added based on convex hull distances, by $W = (1 + e^{-\frac{D}{k T}}) / (1 - e^{-\frac{D}{k T}})$ where $D$ is the distance between the convex hull and the mixing energy of the configuration, $k$, Boltzmann's constant, and $T$ the temperature. After optimization of regularization parameters with respect to k=10-fold cross validation, the test error was \added[id=JMG,comment={Referee 2 comment 1}]{5 meV/site}. The ECIs as a function of cluster radius and numbers of vertices are given in Fig.~\ref{fig:ecis}, with the exception of the identity and single site clusters with ECIs of \added[id=JMG,comment={Retrained cluster expansion with smaller set of parameters to reduce overfitting}]{--15.3 and --2.1 meV}, respectively. The relative sizes of the ECIs reflect the relative contribution to the mixing energy for some surface alloy configuration. Relatively large ECIs associated with clusters containing four vertices highlight the importance of interactions beyond pairs in this system.

\subsubsection{Monte Carlo sampling}
Using stochastic algorithms, a representative 10 $\times$ 10 (2.7 nm) surface cell was sampled in the canonical ensemble at 473K. Only shells with low concentrations of Pd (1-20\% surface content) were considered because of the mechanical destabilization of shells with high palladium content. For each composition, an order parameter was obtained for the chemical ordering motif associated with the the channel shape, a four-point vertically oriented cluster in Fig.~\ref{fig:structure}B along with relevant pair orderings. The relatively accurate predictions of the multi-point orderings were facilitated by sampling converged simulations. The convergence of the parallel Monte Carlo simulations was quantified with the potential scale reduction factor from Brooks and Gelman 1997, which uses the the effective number of independent measurements, $M_{\rm{eff}}(k) = 1/[1+2\lambda (k)]$, and ratios of pooled and within-simulation variances to quantify the convergence of the parallel chains.\cite{brooks_general_1998,gelman_inference_1992} Each parallel simulation was ran for 1000 passes after a 100 pass burn-in. The resulting potential scale reduction factors are on the order of 1.00018 and the simulations are considered well converged.\cite{brooks_general_1998} The statistical analysis and error measurements are detailed further in the Supporting Information.

The MCHammer Monte Carlo software was used through ICET to carry out Markov-chain Monte Carlo samplings of the surface alloy system.\cite{foreman-mackey_emcee_2013,angqvist_icet_2019} The \added[id=JMG,comment={Referee 1 comment 3}]{ClstOPs} were evaluated from the Monte Carlo trajectories using the {\sc Clst\_Order} software package developed for this work. The {\sc Clst\_Order} python software calculates the normalized, symmetry-averaged probabilities in Eq. \eqref{bfunc_prob} for arbitrary cluster shape and order. The parameters can be calculated in general two or three-dimensional crystal systems provided that the atoms can be projected onto pristine parent lattice(s). This software is compatible with the trajectories produced from MCHammer, but can also used with other software packages that generate trajectories or structure files compatible with Atomistic Simulation Environment (ASE) such as the Large-scale Atomic/Molecular Massively Parallel Simulator (LAMMPS) package.\cite{larsen_atomic_2017,plimpton_fast_1995} 

\subsubsection{Chemical ordering quantification}

\begin{figure}[h]
    \centering
    \includegraphics[width=0.5\textwidth]{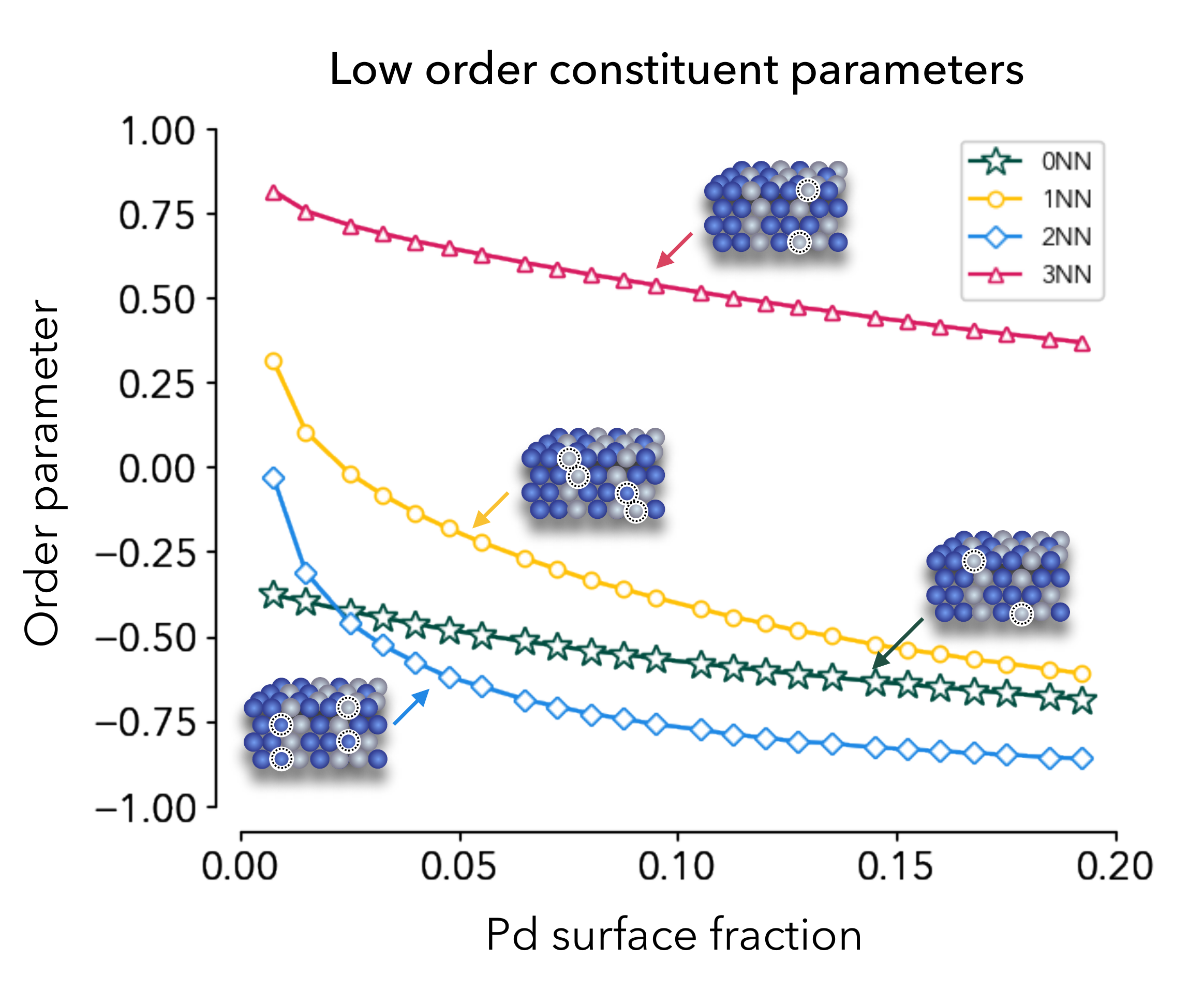}
    \caption{ The pair parameters for all of the nearest neighbor shells contained in the 4-point channel are provided along with the combined single site correlations at the surface and bottom of the alloy.}
    \label{fig:pt_pd_pairs}
\end{figure}

Chemical ordering analysis for alloys is commonly given in terms of pair ordering in certain nearest neighbor shells. Discussions of multi-point ordering motifs generally involve a considerations of all constituent pair orderings contained in the motif.\cite{sauvage_prediction_1974,clapp_theoretical_1969} In the case of the Pd channel, the Pd pair ordering in the 0$^{\rm{th}}$ -- 3$^{\rm{rd}}$ nearest neighbor shells contained in the channel show some limiting factors for the occurrences of the channels as well as the structure of the alloy overall. Due to the anisotropy of the system along the surface norm, pair orderings in a given neighbor shell are not the same throughout the surface alloy; pairs ordering oriented along the surface norm differs from pair ordering parallel to the surface. {For this reason, we extract from the Monte Carlo simulations both sets, the set of 0$^{\rm{th}}$ -- 3$^{\rm{rd}}$ neighbor pair parameters parameters starting from the top of the surface alloy and also from the bottom, of vertically-oriented pair order parameters that can be used for a Kirkwood superposition of the Pd channel. The respective anisotropic consituent pair parameters are reported in Fig.~S3 of the supporting information, while the combined order parameters for the Pd-Pd pairs are reported in Fig.~\ref{fig:pt_pd_pairs} as a function of Pd fraction. The respective combined pair parameters are highlighted in the plot.} In crystals possessing 3D periodicity, this same model can be applied with averaging pair probabilities over all equivalent orientations and constant single-site correlations that are given simply by the concentrations.
The parameters in \added[id=JMG,comment={New results with updated cluster expansion. Pair parameters are now constructed from the pair probabilities starting from the top as pictured in the original figure, but also from the bottom.}]{Fig}.~\ref{fig:pt_pd_pairs} describe how likely Pd-Pd pairs in the first-third neighbor shell of the channel shape relative to a completely random alloy (along with the single site Pd ordering in the zeroth neighbor shell). Recalling that these parameters are zero in a completely random alloy, the lower values for the Pd-Pd pairs in the first neighbor shell (circle markers) indicate that Pd-Pd pairs are likely to form at the top or bottom of the surface alloy. The likelihood for the occurrence precursors of the Pd channels is high. In the third shell (triangle markers), the occurrence of Pd-Pd pairs is highly unlikely, and this is one of the key factors that limits the occurrence of Pd channels overall. The Pd tends to reside in the middle of the shell as indicated by the third neighbor shell parameter as well as the single site correlation for the top/bottom of the alloy. This supports the experimental findings of Zhang et al., because a significant amount of Pd remains after etching of the core. This low order analysis that is common in the literature is useful for determining limiting factors for channel occurrence and alloy structure.
\begin{figure}[h]
    \centering
    \includegraphics[width=0.5\textwidth]{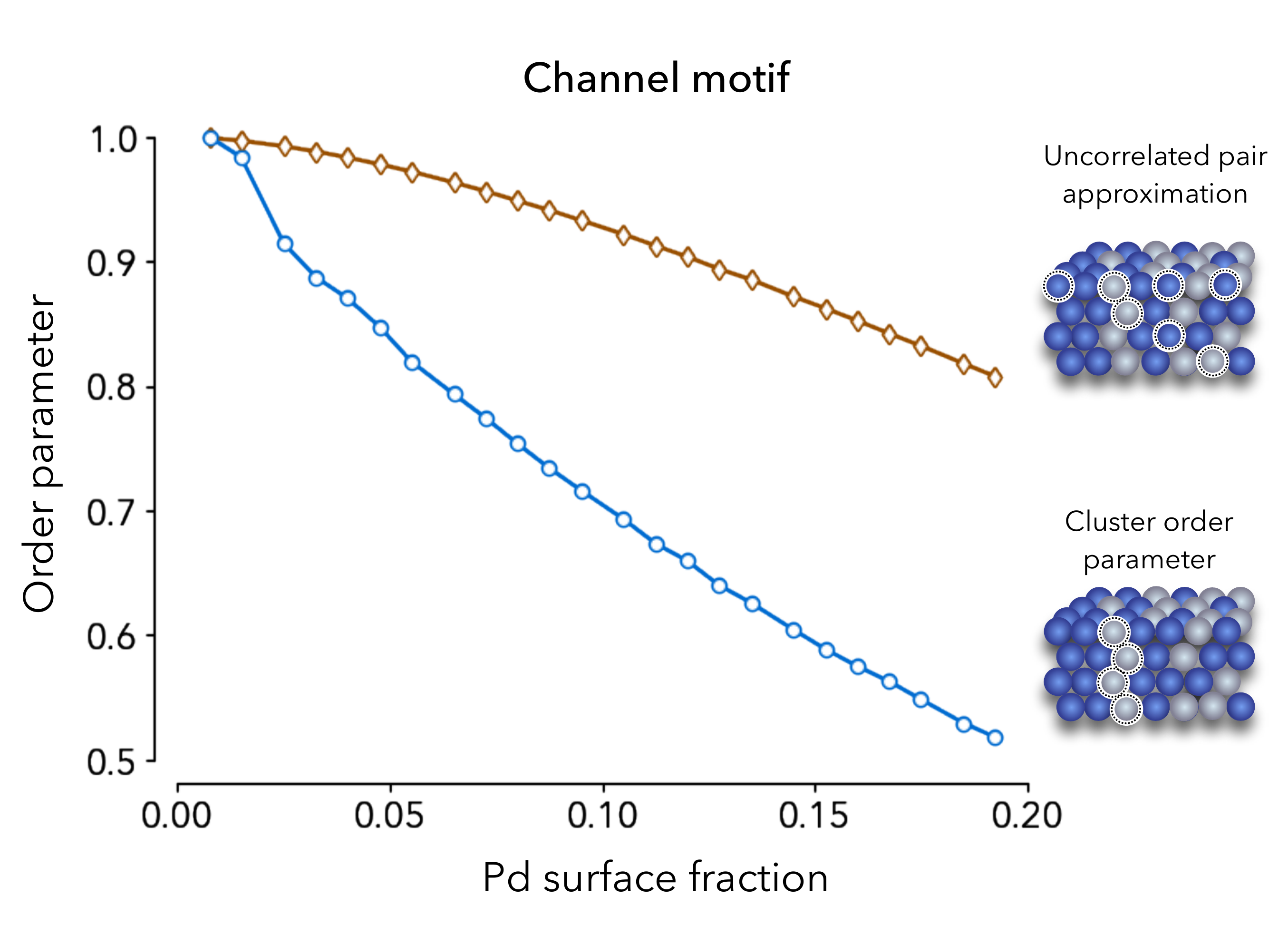}
    \caption{ The exact 4-point order parameter corresponding to the channel-shaped cluster, from direct calculation during MC simulations (blue circles) are compared to that calculated from Clapp's implementation of the Kirkwood superposition (brown diamonds). A visualization of low order probabilities used to approximate the 4-point channel probability are illustrated for each case. }
    \label{fig:pt_pd}
\end{figure}
The Pd channels were quantified during parallel Monte Carlo simulations of the surface alloy using \added[id=JMG,comment={Referee 1 comment 3}]{ClstOPs}, and is reported in Fig.~\ref{fig:pt_pd}. For the compositions tested the channels occur less frequently than in a randomized alloy lattice, because the results show that the \added[id=JMG,comment={Referee 1 comment 3}]{4-point Pd channel ClstOP} is greater than 0. The channels occur only in small amounts. This supports the need for small amounts of Pd channels without excess that mechanically destabilizes the shell in experiment. From pair analysis alone one may expect insignificant amounts of Pd channels at experimentally relevant compositions (~15 atomic \% Pd) given the limiting formation of third neighbor shell Pd-Pd pairs, Fig.~\ref{fig:pt_pd_pairs}, but there are significant amounts of channels when quantified \added[id=JMG,comment={Referee 1 comment 3}]{exactly with ClstOPs}. Using the limiting pair parameter alone is not sufficient for quantifying the Pd channel occurrence. 

Improved quantification of Pd channels can be made by approximating the four-point probability with combinations of pair probabilities. There are various combination/superposition expressions that could be used to approximate the four-point Pd channel probability, but two key examples are provided for comparison. Using a slightly modified form of Eq. 7 in Ref.~\cite{clapp_exact_1967}, the relative four-point probability can be approximated using Kirkwood's superposition,
\begin{equation}
    \bar{P}(\vec{\sigma}_\alpha)/P_{\rm{rand.}} \approx \big( P_0 * P_{01} * P_{02} * P_{03} + \mathscr{P}_{\rm{r}}\big)/{c_{Pd}}^7,
    \label{approx}
\end{equation}
where $P_{ij}$ are the Pd probabilities between the $i^{\rm th}$ and $j^{\rm th}$ neighbor shells and ${c_{Pd}}^7$ is product of the marginal probabilities for the constituent pairs and single site correlations. \added[id=JMG,comment={Explicitly includes other equivalent superpositions in the expression.}]{The $\mathscr{P}_{\rm{r}}$ is the product of the equivalent pair probabilities in reversed order (e.g. starting from the bottom of the surface alloy rather than the top).} The corresponding approximation to the ClstOP is given as the curve with diamond markers in Fig.~\ref{fig:pt_pd}. This approximation describes some qualitative trends correctly, but deviates from the exact ClstOP value. From a consideration of constituent pair probabilities in this way, the amount of Pd channels at low concentrations may still be misleadingly small. One benefit of this approximation is that it can often be obtained from experimentally determined Warren Cowley parameters. It is additionally pointed out that the parameter estimated with cluster probabilities constructed via the Kirkwood superposition are known with less certainty. The graph below shows the relative standard errors in the measured cluster probabilities using the Kirkwood superposition as well as the ClstOP. 
\begin{figure}[h]
    \centering
    \includegraphics[width=0.5\textwidth]{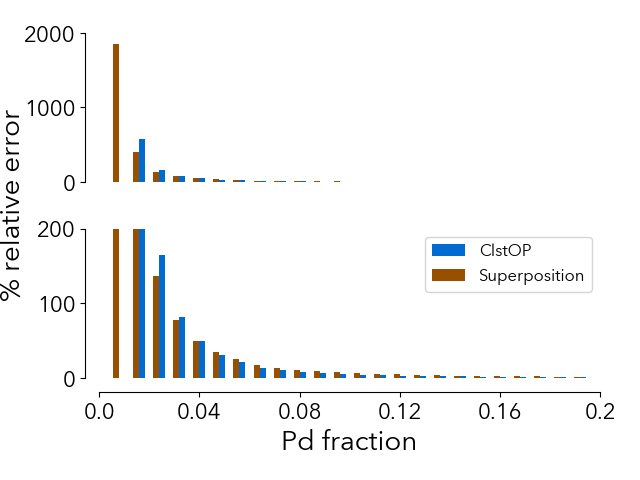}
    \caption{ The relative error in the measured cluster probabilities using the cluster order parameter formalism compared to a constituent pair superposition approximation.}
    \label{fig:pt_pd_err}
\end{figure}

\added[id=JMG,comment={Standard errors for the new CE model along with corresponding discussions of possible deficiencies of the CE model.}]{In Fig.~\ref{fig:pt_pd_err}, it is shown that the relative standard error for the measured quantities for the ClstOP ranges between 0.9 \% and  580.1 \% over the full Pt composition range tested ( 0.9 - 21.7\%  over the experimentally relevant composition range of $c_{\rm{Pd}} =$ 0.05 - 0.2 ). The Kirkwood superposition gives a relative standard error range of  1.6 - 1843.6 \% over the full composition range tested ( 1.6 - 25.1 \% over the experimentally relevant composition range). After propagation of error, the approximation method quantifies the Pd channel occurrence with larger uncertainty than the exact ClstOP due to the multiple measurements needed to construct the Kirkwood superposition. The increased relative error in the low Pd limit is due to the low Pd concentration, which is related exponentially to the probability of observing a Pd channel, and possibly the quality of the model in the very dilute Pd limit for the cluster expansion model (See Supporting information Fig. S2).}

The approximation suggested by Shirley and Wilson in 1977 may provide a slightly improved prediction of the exact parameter in the surface alloy system. For this case, the four-point probability is approximated as,
\begin{equation}
    \bar{P}(\vec{\sigma}_\alpha) \approx \big( P_{01} \times P_{23} + P_{02} \times P_{13} + P_{03} \times P_{12} \big),
    \label{shirley}
\end{equation}
This approximation is not given within because it relies on spatially resolved ordering probabilities that are not obtained from typical experimentally derived pair parameters. In many 3D alloy crystals, the cluster probabilities are averaged over the distinct cluster locations needed to construct the 4-point probability in Eq. \eqref{shirley}. 

\section{Conclusion}
A general cluster order parameter (ClstOP) was introduced to systematically quantify multi-point ordering motifs in alloy crystals via direct calculation of cluster probabilities. This parameter can be used to quantify chemical ordering in alloys and substitutional systems that cannot be addressed by pair ordering analysis alone. Though the pair ordering and pair interactions are often most important, there are systems where higher order correlations are significant and cannot be ignored. The utility of the \added[id=JMG,comment={Referee 1 comment 3}]{ClstOPs} is that a specific multi-point motif of interest can be quantified directly in simulations or in theoretical applications. In certain special cases where 'clusters' are chosen such that they can represent derivative structures of a crystal, these parameters could be used to help generalize the geometrical locus method. 

Despite the expected low probability of occurrence for multi-point ordering motifs, meaningful predictions of \added[id=JMG,comment={Referee 1 comment 3}]{ClstOPs} can be made through efficient sampling during parallel Markov-chain Monte Carlo simulations. The average \added[id=JMG,comment={Referee 1 comment 3}]{ClstOPs} can be predicted with reasonable certainty with relative standard errors of 21.7 - 0.9 \%  between the experimentally relevant composition range of $c_{\rm{Pd}} =$ 0.05 - 0.20, respectively. The \added[id=JMG,comment={Referee 1 comment 3}]{ClstOP associated with the 4-point Pd channel motif} quantifies Pd channel occurrence with improved certainty and accuracy over approximate methods such as the Kirkwood superposition. Similar sampling approaches could be used to predict multi-point ClstOPs in many other practical applications such as descriptors for data-driven materials discovery and machine-learning models of alloy systems.\cite{jennings_genetic_2019} With the multi-lattice generalization, \added[id=JMG,comment={Referee 1 comment 3}]{chemical orderings} between multiple sublattices can also be described. This generalization is particularly useful for describing ordering between ligand vacancies on one sublattice and alloying metals in another.

The utility of the parameters was demonstrated while modeling representative surfaces of the Pt/Pd nanoparticle alloy system in Fig.~\ref{fig:structure}. Cluster expansion models were generated for a four-layer dilute Pt/Pd alloy on top of a four-layer bulk palladium core region with the (111) surface orientation, and these systems were sampled using Markov-Chain Monte Carlo simulations. In this system, Pd channels that span the surface alloy are needed for the synthesis of high surface area Pt nanocage catalysts. \added[id=JMG,comment={Predicted channels per octahedral nanoparticle are now reported quantitatively.}]{The calculated ClstOP corresponding to a channel shape in these simulations suggested that Pd channels occur in significant amounts, with 0.12 per nanoparticle with experimental facet sizes (octahedra with 19.4 ${\rm{\AA}}$ edge lengths at a 15 atomic \% Pd in the surface). Using pair probabilities alone to quantify Pd channels leads to a significant underestimation of channel ordering, and less than half as many channels occurring with a predicted 0.05 per nanoparticle.} Though pair ordering provides a wealth of information about the alloy structure overall and the factors that limit the occurrence of a Pd channel, it provides poor estimates of the true number of channels occurring in the alloy. This is because multi-point energetics are significant in this system. Additional results in the Supporting Information suggest that different solvent environments could aid in the design of nanostructured catalysts that are more durable. Solvation induces a surface enrichment of Pd without completely eliminating the occurrence of Pd channels that span the surface alloy. This was shown by exact quantification of the four-point channel motif where the limiting factors derived from pair ordering alone may lead one to believe that the Pd channels do not occur in significant amounts.

\section{Data availability}
The data is available upon reasonable request from the authors. Software to calculate the order parameters is available at https://github.com/jmgoff/clst\_order.

\section{Acknowledgements}
J. M. G., S. B. S., and I. D. acknowledge financial support from the U.S. Department of Energy, Office of Science, Basic Energy Sciences, CPIMS Program, under Award No. DE-SC0018646. \\
First-principles calculations for this research were performed on the Pennsylvania State University's ROAR supercomputer.\\
A portion of the Monte Carlo analysis was performed on the CNMS CADES supercomputer at Oak Ridge National Laboratory

\clearpage
\balance
\bibliography{order_parameters}
\end{document}


\title{Quantifying multi-point ordering in alloys: Supporting information}

\author{James M. Goff}
\email{jmg670@psu.edu}
\affiliation{Department of Materials Science and Engineering, The Pennsylvania State University, University Park, PA 16802, USA}
\author{Bryant Y. Li}
\affiliation{Department of Materials Science and Engineering, The Pennsylvania State University, University Park, PA 16802, USA}
\author{Susan B. Sinnott}
\affiliation{Department of Materials Science and Engineering, The Pennsylvania State University, University Park, PA 16802, USA}
\affiliation{Department of Chemistry, The Pennsylvania State University University Park PA 16802, USA}
\affiliation{Materials Research Institute, The Pennsylvania State University, University Park, PA 16802, USA}
\author{Ismaila Dabo}
\affiliation{Department of Materials Science and Engineering, The Pennsylvania State University, University Park, PA 16802, USA}
\affiliation{Materials Research Institute, The Pennsylvania State University, University Park, PA 16802, USA}
\affiliation{Penn State Institutes of Energy and the Environment, The Pennsylvania State University University Park PA 16802, USA}
\maketitle

\section{Statistical sampling methods}

A parallel sampling of $S$ = 100 simulations was used to obtain the globally averaged order parameters in Fig. 5 of the main text as the average of ensemble means from each simulation, $s$.
\begin{equation}
    \bar{\gamma} = \frac{1}{S}\sum_s^S \frac{k}{M} \sum_{mk}^M \gamma_{mk},
    \label{global_avg}
\end{equation}
The average of each simulation, $s$, is calculated in the inner sum. The order parameter at a microstate, $m$, at a subsampling frequency, $k$. E.g. the microstates in the Markov chain are sampled every $k$ steps to evaluate the average. A thorough sampling such as this is needed to obtain sufficient statistics for the cluster order parameters; the random probability for a Pd channel to occur in the surface alloy for the lower Pd concentrations tested are around $1:10^8$. Attempting to sample this many microstates at a sampling frequency of $k$ = 1 may be computationally intensive, especially in the regime that time to measure the quantity is large compared to the time to progress the Markov-Chain by one step. Instead, subsampling a chain at some frequency $k>1$ may lead to an increase in statistical efficiency or reduce excessive autocorrelations between measurements if present.\cite{owen_statistically_2017} The relative efficiency of subsampling at frequency $k$ compared to $k$=1 from Owen 2017 is,
\begin{equation}
    eff(k) = \bigg(\frac{1+\rho}{k+\rho}\bigg) \bigg(\frac{1+2 \sum_m r_m}{1+2 \sum_m r_{km}}\bigg),
    \label{eff}
\end{equation}
where $\rho$ is the ratio of the time taken to calculate the order parameter over the time to progress the Markov Chain by one step, which is 9.5 with the current software used. The efficiency depends on the sum of the autocorrelation function, $r$, of the measured order parameter. By maximizing Eq. \eqref{eff} with respect to $k$, it is found that a subsampling an optimized frequency of $k=10$ results in a 4.82-fold efficiency gain when measuring the 4-point order parameter in the Pt/Pd surface alloy. Each simulation in Eq. \eqref{global_avg} is sampled at this frequency to improve the efficiency. An added benefit for subsampling is that reduces the number of correlated measurements of the order parameter. 

The relative errors reported in Fig. 3 of the main text account for the effective number of measurements and are estimated as,
\begin{equation}
    {\rm{std. \, err.}} \, = \, \sqrt{ \frac{\varsigma^2}{M_{\rm{eff}}S} \; + \; \frac{\varsigma_{\rm{s}}^2}{S}}
    \label{std_err}
\end{equation}
where $\varsigma$ is the variance of the order parameter in the individual simulations, $\varsigma_{\rm{s}}$ is the spread of the posterior mean order parameters across the $S=100$ simulations, and $M_{\rm{eff}}$ is the effective number of independent measurements. Though the error in the DFT and cluster expansion energies are not accounted for, these results and analysis have shown that statistically meaningful predictions of multi-point ordering can be made through a combination of parallel sampling and consideration of autocorrelation.

\section{Note on the ClstOP response relative to the random alloy phase}
\added[id=JMG,comment={Section added regarding the fluctuations of the random alloy phase, and how the cluster order parameter responses can generally be distinguished from random alloy fluctuations except in very specific cases.}]{The ClstOPs}, much like the Warren-Cowley pair parameters, describe chemical ordering relative to the random alloy phase. In the random alloy phase, the cluster correlations by definition are all 0, the variance however is not zero. One may be concerned that the measured response is not significant, given random alloy fluctuations of the cluster probabilities. The accuracy and limitations of the ClstOPs are discussed here bearing this in mind. To see whether or not the measured cluster probability (and therefore the measured ClstOP) is distinguishable from random alloy fluctuations, a two-sample t-test can be performed. Provided that the t statistic evaluates to a value larger than the critical t-value for some confidence interval, then it can be confirmed that the cluster probability associated with the ClstOP isn't just a random alloy fluctuation at that confidence interval.
\begin{equation}
    t = \frac{(\bar{P}_\alpha^{measure} - \bar{P}_\alpha^{random})}{\sqrt{{\rm{var}^{pool}}(P_\alpha)*\bigg(  \frac{1}{n_{measure}} + \frac{1}{n_{random}} \bigg)}}
    \label{ttest}
\end{equation}
where $n_{measure}$ and $n_{random}$ are the number of samples taken in the measured and random alloy, respectively. The pooled variance is ${\rm{var}^{pool}}(P_\alpha)$. To obtain this pooled variance, the variance of both the measured cluster probability and that in the random alloy need to be known. The variance of the measured cluster probability is given during the calculation of the cluster correlations, but the variance of the probability in the random alloy phase needs to be defined. 

First, the variance of the cluster correlations in a single structure is derived. This variance is given as,
\begin{equation}
    {\rm{var}}(\bar{\Phi}_\alpha) = \frac{1}{Nm_\alpha}\big[\bar{\Phi^2}_\alpha - \big(\bar{\Phi}_\alpha \big)^2 \big]
    \label{corr_var}
\end{equation}
where the case of vanishing cluster correlations. In an \textit{N}-site, equimolar random alloy, Eq. \eqref{corr_var} becomes:
\begin{equation}
    {\rm{var}}(\bar{\Phi}_\alpha) = \frac{1}{Nm_\alpha}\sum_{\vec{\sigma}_\alpha} \Phi^2_\alpha(\vec{\sigma}_\alpha) \bar{P}(\vec{\sigma}_\alpha)
    \label{random_var1}
\end{equation}
where the cluster correlation function squared is measured in one structure. The cluster probabilities in the random alloy are given in terms of the marginal probabilities of the constituent lattice sites, $P_{random} = \prod _{i \in \alpha} c_i$ where the product runs over all elements $i$ that occupy the cluster sites, and $c_i$ is the concentration of the species at site $i$. Using this, the variance can be further simplified.
\begin{equation}
    {\rm{var}}(\bar{\Phi}_\alpha) = \frac{1}{N m_\alpha}\sum_{\vec{\sigma}_\alpha} \Phi^2_\alpha(\vec{\sigma}_\alpha) \prod_{i \in \alpha} c_i
    \label{random_var2}
\end{equation}
Other than variance decreasing with increasing system size, $N$ in Eq. \eqref{random_var2}, it is not very clear how the equation scales for a given cluster. For a cluster with $\nu$ vertices on a single sublattice, there are $d^\nu$ possible cluster occupations where $d$ is the number of occupational degrees of freedom. Because $\Phi_\alpha \in [-1,1]$, and $\Phi_\alpha^2 \in [0,1]$, the first summed term scales roughly as $d^\nu$ and by a constant factor, determined by site basis normalization, that depends only on the degrees of freedom and the number of vertices in the cluster. The second summed term scales roughly as $c^\nu$, where $c$ is the average concentration of alloying species. The variance therefore scales roughly as,
\begin{equation}
    {\rm{var}}(\bar{\Phi}_\alpha) \approx \frac{1}{N m_\alpha} f(d,\nu) d^\nu c^\nu = \frac{f(d,\nu)}{N m_\alpha} 
    \label{approx_var2}
\end{equation}
where the $d^\nu c^\nu$ term equals one. This shows the intuitive scaling found by Zunger et al. in 1990, but is generalized to multiple components.\cite{zunger_special_1990}

Because there are equimolar alloying species in this case, $c_A = c_B = c_C ... c_i$, Eq. \eqref{random_var2} can be further simplified. For all possible cluster occupations, the product inside the sum is the same, $\prod _{i \in \alpha} c_i =c_i ^\nu$, and it can be removed from the sum. The variance of the cluster correlation becomes,
\begin{equation}
{\rm{var}}(\bar{\Phi}_\alpha) = \frac{1}{N m_\alpha}c_i^\nu \sum_{\vec{\sigma}_\alpha} \Phi^2_\alpha(\vec{\sigma}_\alpha),
    \label{equimolar_var}
\end{equation}
and simplifies exactly to the right-most expression in Eq. \eqref{approx_var2}. For the case of binary alloys, the expression for the variance reduces exactly to what is found in Zunger 1990, ${\rm{var}}(\bar{\Phi}_\alpha) = \frac{1}{N m_\alpha}$.

In a 5-component FCC-based crystal, the t-statistic is checked using Eq. \eqref{equimolar_var}. The NN pair and NN triplet cluster probabilities are measured, in a symmetrically distinct 5-site cell where the variance is largest. By taking the degeneracy of the structures into account as multiple measurements of the cluster occupations and calculating the t-statistics, the confidence in the measurements in table \ref{tab:tscores} are obtained.
\begin{table*}
    \caption[Confidence intervals for ClstOP measurements]{Confidence intervals and number of ClstOP measurements in the 5-component FCC-based crystal. }
    \centering
    \begin{tabular*}{0.9\textwidth}{@{\extracolsep{\fill}}ccccc}
        \hline
        \hline
        \\
        Dataset & \multicolumn{4}{c}{Measurements of ClstOP at confidence intervals}
        \\
        \\
        \hline
        \\
        \\
        CI & 80-90\% & 90-95\% & $>$=99\% & $>$99.9\% \\
        \\
        N=5 \\
        \\
        \multicolumn{1}{r} \# of measurements & 5/25  & 0/25 & 0/25 & 20/25 \\
        \\
        \hline
        \\
        N=25 \\
        \\
        \multicolumn{1}{r} \# of measurements & 0/125 & 0/125 & 0/125 & 125/125 \\ \\
        \\
        \hline
        \hline
        \\
    \end{tabular*}
    \label{tab:tscores}
\end{table*}
The confidence in the ClstOP measurements reported in Table \ref{tab:tscores} shows that the ClstOP responses are known with significant certainty relative to the random alloy phase. These certainties are likely to increase in larger cells by Eq. \eqref{equimolar_var}. Additionally, the certainty is likely to increase for larger clusters because the random alloy phase variance decreases with cluster size. 

In the Pt/Pd surface alloy system, the chemical degrees of freedom are lower and the variance in Eq. \eqref{random_var2} decays more slowly than for the equimolar 5-component example. If one were to measure the ClstOP response in a very small unit cell of a binary system, the response may be insignificant. The cluster probabilities in the Pt/Pd surface alloy system are obtained by averaging over many states belonging to the canonical ensemble, greatly increasing the number of samples, $n_{measure}$, in the t-statistic. The t-statistic for the composition with the largest relative variance of the cluster probability (worst case for the surface alloy measurements), is -282. This means that the cluster probability is distinguishable from a the random alloy probability with a certainty well beyond the 99.9 \% confidence interval. In general, it may not be necessary to consider the fluctuations of the random alloy phase to assess the certainty of the measured ClstOP response. This is especially true for responses measured as an ensemble average during simulations. For these cases, it will be more important to ensure that the simulations are well converged with respect to the autocorrelation of the measured ClstOP response. Care should be taken with measuring the ClstOP response in very small cells of binary/ternary systems (N $\approx$ $d$), as seen in Table S\ref{tab:tscores}. In these cases the ClstOP response may be negligible compared to random alloy fluctuations. For small unit cells such as these, it will likely be more useful to quantify long-range order.

\section{Supporting solvation and statistical sampling}

\begin{figure}[h]
    \centering
    \includegraphics[width=0.85\textwidth]{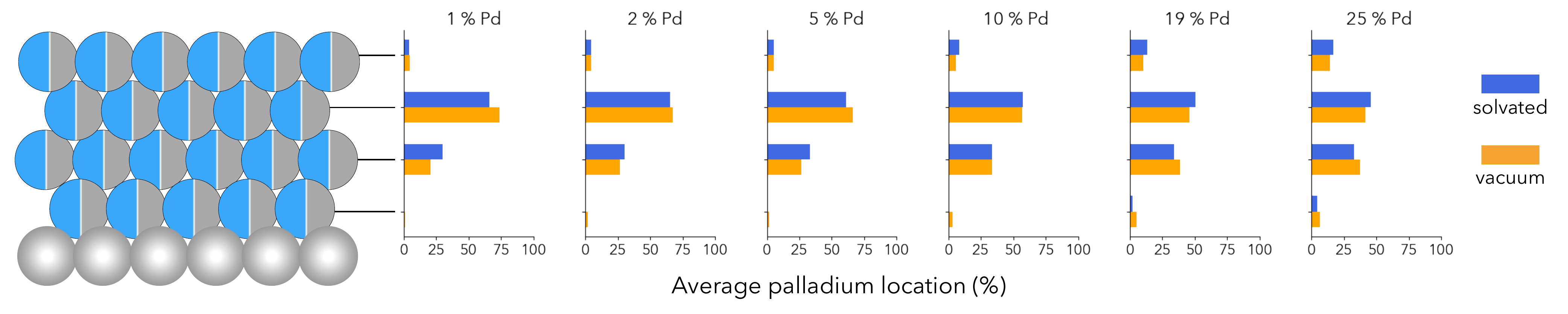}
    \caption{Average Pd location in the surface alloy in vacuum (orange) and solvent (blue) during MC simulations. The concentration of Pd in upper layers of the surface alloy is increased in the presence of the solvent.}
    \label{fig:bar_pic}
\end{figure}

When the simulations and analysis are repeated for a surface alloy in vacuum, the trends in chemical ordering are similar. Channels are less likely to occur in canonical microstates than in a randomized alloy lattice at the same concentration. An etching step for the synthesis of the nanocatalysts does not drastically change the number of palladium channels in the system, but the solvent does induce an enrichment of palladium atoms closer to the surface, as seen in Fig. S\ref{fig:bar_pic}. The palladium atoms on the surface of the alloy would be removed during etching. This suggests that solvent treatment techniques could be used to drive palladium towards the nanocatalyst surfaces to obtain platinum shells that are more mechanically stable/resistant to corrosion.

\begin{figure}[H]
\centering
\begin{subfigure}{.5\textwidth}
  \centering
  \includegraphics[width=1.\linewidth]{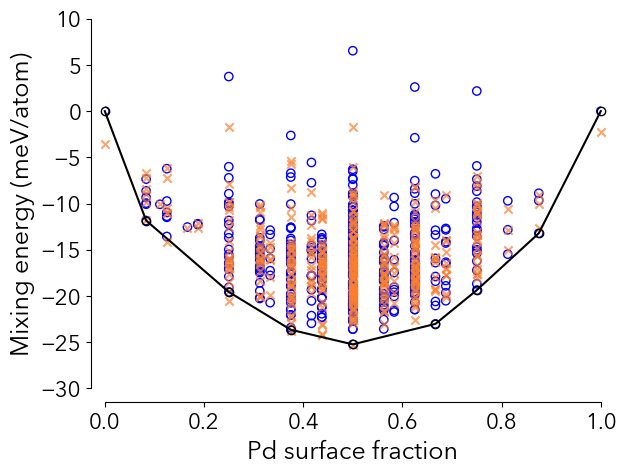}
  \label{fig:sub1}
\end{subfigure}%
\begin{subfigure}{.5\textwidth}
  \centering
  \includegraphics[width=1.\linewidth]{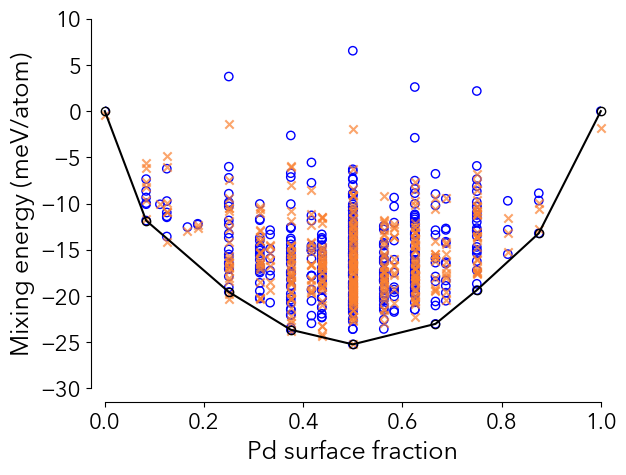}
  \label{fig:sub2}
\end{subfigure}
\caption[Mixing energy hull of the Pt/Pd surface alloy]{(left) Convex hull of mixing enthalpy for the Pt/Pd surface alloy system in the top four layers. The DFT calculated energies are given as blue circles, the cluster expansion predictions as orange 'x's and the convex hull is marked with a black line. The composition on the x axis is the Pd fraction of the topmost 4-layer region of the cell. (right) An overconstrained fit of the DFT convexhull leads to better predictions of end member energies, but poorer predictions of mixed-composition energetics.}
\label{fig:convex}
\end{figure}

The DFT-predicted mixing energies of the platinum palladium surface alloy are compiled in the convex hull in Fig.~S\ref{fig:convex}. It is noted that the DFT-calculated energies of cells containing palladium channels are higher in energy compared to other alloy configurations where palladium occupies the sites in the middle of the alloy shell. \added[id=JMG,comment={More on the topic of overfitting: unconstrained vs constrained convex hulls.}]{In Fig.~S\ref{fig:convex} (left), the endpoints are predicted poorly compared to the rest of the datapoints. This is likely due to the calculation of ECIs with many more structures with compositions deviating further from the end member compositions. This is commonly addressed by applying constraints on predicted energies of the mixing energy end members but applying constraints to the predicted energies of some structures may lead to poor predictions of the energies of other structures. This may be observed in Fig.~S\ref{fig:convex} (right) where the end members are predicted more accurately there are more poorly predicted mixing energies in the 10-20 atomic \% Pd region that is of interest. For this reason, the end member fitted energies are not constrained. Instead of constraining fitted mixing energies, improved models may instead add chemical or lattice-parameter dependence to the ECIs.\cite{sanchez_foundations_2017,cao_computationally_2019} The less accurate predictions of the end members may lead to poor predictions of alloy structure and stability in the very low and very high Pd concentration regimes, but are expected to be more accurately predicted in the 10-20 atomic \% Pd composition range. }

\section{Reduction of the cluster order parameter to Warren-Cowley parameters}
It can easily be shown that the cluster order parameter reduces to the Warren-Cowley parameter for pair clusters on a single sublattice. First we define the cluster that we are interested in as a pair, $\alpha \; \in \; \{ pairs \}$ and that the occupation of the pair consists of two different species $p$ and $q$ giving the desired occupation vector $\vec{\sigma}_\alpha^d =
\begin{bmatrix}
\sigma_p,\\
\sigma_q\\
\end{bmatrix}$. When we evaluate the cluster order parameter for this occupation, we obtain:
\begin{equation}
    \gamma_\alpha =\;\; 1 \; - \; \frac{\bar{P}(\vec{\sigma}_\alpha^{d})}{P(p)\,P(q)}
    \label{pair_clst}
\end{equation}
and the average probability in the numerator can be interpreted as the measured probability of both $p$ and $q$ being in the pair cluster, $P(p \; and \; q)$ since it is averaged over all symmetric permutations of the pair. Making note of this and rewriting the pair cluster order parameter in Eq. \eqref{pair_clst}, we obtain the ratio between the joint probability of both $p$ and $q$ being present in the pair and the individual probabilities of finding $p$ or $q$ in the crystal. Plugged in on the left-hand side along with the individual probabilities in the denominator,
    \begin{equation}
        \gamma_\alpha = 1\;-\;\frac{P(p \; \cap \; q)}{c_p c_q} \; = \; 1 \; - \frac{P(q|p)}{c_q},
    \end{equation}
shows that the cluster order parameter is equivalent to the binary Warren-Cowley parameter using the relationship between joint and conditional probabilities. The probability of finding a certain species $p$ or $q$ on a site in a random alloy is given by the respective concentrations, $c_p$ and $c_q$. The specific nearest neighbor shell that one would specify in the Warren-Cowley parameter, $m$, is now just specified by the cluster itself, $\alpha$. It is noted that the cluster order parameters are not independent from other order parameters; there are constraints based on the composition of the substitutional system. Therefore, cluster order parameters for pairs do not reduce to the multicomponent, linearly independent pair parameters defined in de Fontaine 1971.\cite{fontaine_number_1971}

\section{Extra information for pair superposition approximation}
\begin{figure}[h]
    \centering
    \includegraphics[width=\textwidth]{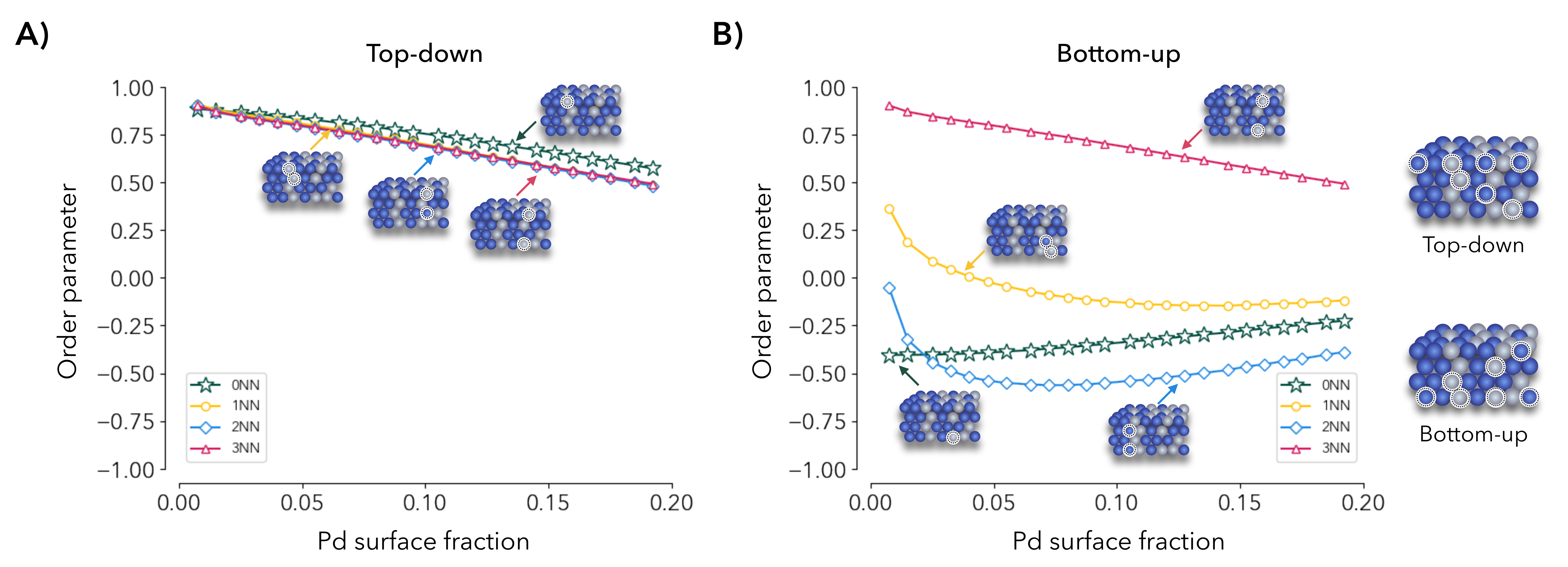}
    \caption{ A) The pair parameters for all of the nearest neighbor shells contained in the 4-point channel are provided along with the single site Pd correlation starting from the top of the surface alloy. B) The same as in A) but from the bottom of the surface alloy going up. The corresponding combinations are shown on the right.}
    \label{fig:all_pairs}
\end{figure}

\added[id=JMG,comment={Expanded discussion on the limiting behavior to justify results with the refined CE model}]{A} superposition of pair probabilities to construct the 4-point Pd channel may be performed in multiple ways for this system: starting from the top of the surface alloy and moving down, start from the center of the surface alloy and moving out, etc. Starting from the center requires pair probabilities zeroth, first, and second neighbor shells while starting from the surface uses zeroth, first, second, and third neighbor shells. Starting from the center, however, requires the product of two separate first NN pair probabilities. Possible superposition combinations such as this with products of parameters in the same neighbor shell were omitted from consideration. They require spatially resolved chemical ordering data that is not possible to get with experimental data, and would be superposition approximations of a different order in a 3D crystalline lattice. Instead, the superposition combinations are constructed through products of pair probabilities in the zeroth, first, second, and third neighbor shells, starting from the top of the shell, and moving down. Equivalently, a superposition \textit{of the same order} may be constructed by taking a product of pair probabilities in the zeroth, first, second, and third neighbor shells starting from the bottom of the surface alloy region and moving up. The constituent pair parameters for each case are reported in Fig. \ref{fig:all_pairs} A and B, respectively.

The pair parameters reported in Fig. 5 of the main text are given from the combined pair probabilities in Fig. S\ref{fig:all_pairs} A and B, and the reported superposition in Fig. 6 of the main text results from the direct addition of the superpositions constructed from the products of the associated pair probabilities for Fig. S\ref{fig:all_pairs} A and B. 

As suggested in the main text, the pair parameters are quite useful for determining limiting behavior. In this surface alloy system with broken symmetry along the surface norm, the pair parameters starting from the top vs the bottom provide additional information that wouldn't be necessary in bulk 3D crystals. That is that the limiting behavior for the Pd-Pd pairs is the Pd surface content, which is significantly lower than what would be expected based on the nominal stoichiometry, Fig. S\ref{fig:all_pairs} A. A significant amount of Pd tends to reside in the bottom and middle two layers of the surface alloy, Fig. S\ref{fig:all_pairs} B. 

\clearpage
\balance
\bibliography{order_parameters}